\documentclass[doublecol,figures]{epl2}
\usepackage{amsmath}

\title{Herding model and \mth{1/f} noise}

\author{J. Ruseckas \and B. Kaulakys \and V. Gontis}
\shortauthor{J. Ruseckas \etal}

\institute{
  Institute of Theoretical Physics and Astronomy, Vilnius University,
  A.~Go\v{s}tauto 12, LT-01108 Vilnius, Lithuania
}
\pacs{05.40.-a}{Fluctuation phenomena, random processes, noise, and Brownian motion}
\pacs{89.65.Gh}{Economics; econophysics, financial markets, business and management}
\pacs{05.10.Gg}{Stochastic analysis methods}

\abstract{
We provide evidence that for some values of the parameters a simple
agent based model, describing herding behavior, yields signals with $1/f$ power
spectral density. We derive a non-linear stochastic differential equation for
the ratio of number of agents and show, that it has the form proposed earlier
for modeling of $1/f^{\beta}$ noise with different exponents ${\beta}$. The
non-linear terms in the transition probabilities, quantifying the herding
behavior, are crucial to the appearance of $1/f$ noise. Thus, the
herding dynamics can be seen as a microscopic explanation of the proposed
non-linear stochastic differential equations generating signals with
$1/f^{\beta}$ spectrum. We also consider the possible feedback of macroscopic
state on microscopic transition probabilities strengthening the non-linearity of
equations and providing more opportunities in the modeling of processes
exhibiting power-law statistics.}

\begin{document}

\maketitle

\section{Introduction}

Kirman's seminal herding model was introduced in
Refs.~\cite{Kirman1991,Kirman1993}. This is a simple stochastic model of
information transmission initially designed to explain the herding behavior in
ant colonies, gathering food from two identical sources located in their
neighborhood. Kirman noticed that entomologists and economists observe similar
patterns in rather different systems. If there are two identical food sources
available to ants in a colony, majority of ants still tend to use only a single
food source at any given time. Furthermore, switches to the new food source
occur despite the fact that food sources remain identical \cite{Detrain2006}.
Human crowd behavior tends to be quite similar, at least in statistical sense.
There are observations that majority of people tend to choose more popular
product, than less popular, despite both being of a similar quality. The article
\cite{Kirman1993} also cites numerous works, which speculate that herding
behavior might be related to the fluctuations of asset price. In Kirman's
formalization the switching probabilities do not depend on the source, thus
probability distribution of ant's visiting times at both mangers is symmetric.

Kirman's model and similar approaches have been applied as models of herding and
contagion phenomena in financial markets \cite{Kirman1991,Kirman2002,Lux2000}.
In ref.~\cite{Constantini1997} the equilibrium distribution of the related
discrete-time stochastic process within the more general theoretical framework
of Polya urn processes has been derived. The associated Fokker-Planck equation
for the pertinent continuous symmetric dynamic process has been derived and
solved in ref.~\cite{Alfarano2008}. The parameters of Kirman's herding model
applied to the description of financial markets were estimated in
ref.~\cite{Gilli2003} by introduction of a simulated moment approach extracting
two key parameters of the model via matching of the empirical kurtosis and the
first autocorrelation coefficient of squared returns. A direct estimation of the
parameters of a related agent-based model, based on a closed-form solution of
the unconditional distribution of returns, has been proposed in
ref.~\cite{Alfarano2005}. The Kirman model was generalized in
ref.~\cite{Aoki2002}. It is worth to notice that the appropriate
agent-based models can yield emergence the power-law scaling, long-range
correlations, (multi)fractality and fat tails (see,
e.g.,~\cite{Lux1999,Passos2011} and references herein), however the omnipresent
$1/f$ noise have not yet been revealed in such approach.

The phrases ``$1/f$ noise'', ``$1/f$ fluctuations'', and ``flicker noise'' refer
to the phenomenon, having the power spectral density at low frequencies $f$ of
signals of the form $S(f)\sim1/f^{\beta}$, with $\beta$ being a system-dependent
parameter. Power-law distributions of spectra of signals with $0.5<\beta<1.5$,
as well as scaling behavior in general, are ubiquitous in physics and in many
other fields, including natural phenomena, human activities, traffics in
computer networks and financial markets
\cite{Scholarpedia2007,Weissman1988,Barabasi1999,Wong2003,
Newman05,Szabo2007,Castellano2009,Perc2010,Orden2010,Kendal2011,Torabi2011}.
This subject has been a hot research topic for many decades (see, e.g., a
bibliographic list of papers by Li \cite{Li2011}, and a short review in
Scholarpedia \cite{Scholarpedia2007}). Despite the numerous models and theories
proposed since its discovery more than 80 years ago
\cite{Johnson1925,Schottky1926}, the subject of $1/f$ noise remains
still open for new discoveries. Most of the models and theories
are not universal because of the assumptions specific to the problem
under consideration. In 1987 Bak \textit{et al} \cite{Bak1987} introduced the notion of
self-organized criticality (SOC) with the motivation to explain the universality
of $1/f$ noise, as well. Although the paper \cite{Bak1987} is the most cited
paper in the field of $1/f$ noise problems, it was shown later on
\cite{Jensen1989,Kertesz1990} that the mechanism proposed in \cite{Bak1987}
results in $1/f^\beta$ fluctuations with $1.5 < \beta < 2$ and does not explain
the omnipresence of 1/f noise. On the other hand, we can point to a recent
paper \cite{Baiesi2006} where an example of $1/f$ noise in the classical
sandpile model has been provided. A short
categorization of the theories and models of $1/f$ noise is presented in the
introduction of the paper \cite{Kaulakys2009}.

Recently, the nonlinear stochastic differential equations (SDEs) generating
signals with $1/f$ noise were obtained in Refs.~\cite{Kaulakys2004,Kaulakys2006}
(see also recent paper \cite{Kaulakys2009}), starting from the point
process model of $1/f$ noise
\cite{Kaulakys1998-1,Kaulakys1999-1,Kaulakys2000-1,Kaulakys2000-2,Gontis2004,Kaulakys2005}.
Nonlinear SDEs provide macroscopic description of a complex system. Microscopic,
agent based reasoning of equations exhibiting $1/f$ noise can yield further insights
into behavior of the system. In this paper we show that it is possible to
obtain nonlinear SDE of the form of Refs.~\cite{Kaulakys2004,Kaulakys2006}
starting from agent-based herding model. Thus, it is possible to show
analytically that nonlinear nature of herding interactions and
appropriate choice of variable results in $1/f$ fluctuations.

\section{The herding model}

\label{sec:herding}In papers \cite{Kirman1991,Kirman1993,Kirman2002}
Kirman employed a simple model to describe the behavior of a multitude
of heterogeneous interacting agents. In the model the dynamic evolution
is described as a Markov chain. There is a fixed number $N$ of agents,
each of them being in state $1$ or in state $2$. The number of agents
in the first state is denoted by $n$, and the number in the second
state by $N-n$. The core of Kirman's model is the pairwise interaction
governing the transition of the agents between the two states. Neither
the probability of following another companion nor the success in
recruiting companions depend on the outcome of previous meetings.
The lack of memory of the agents is the crucial assumption to formalize
the population dynamics as a Markov process. Describing the dynamics
as a jump Markov process in continuous time, the transition probabilities
per unit time are given by
\begin{align}
p(n\rightarrow n+1) & \equiv p^{+}(n)=(N-n)(\sigma_{1}+hn)\,,\label{eq:pp}\\
p(n\rightarrow n-1) & \equiv p^{-}(n)=n(\sigma_{2}+h(N-n))\,.\label{eq:pm}
\end{align}
The above probabilities define a one-step stochastic process
\cite{vanKampen1992}. The constants $\sigma_{1}$ and $\sigma_{2}$ describe the
idiosyncratic propensity to change the state, while the term $h$ describes the
herding tendency. We allow the two constant parameters $\sigma_{1}$ and
$\sigma_{2}$ to assume different values, generating asymmetric behavior. The
non-linearity in the transition probabilities (\ref{eq:pp}) and (\ref{eq:pm})
constitutes a crucial ingredient of the model: the presence of non-linear terms,
is the imprint of interactions among agents. The linear terms would imply
independence of the behavior of the agents.

The transition rates (\ref{eq:pp}) and (\ref{eq:pm}) describe a scenario where
the interaction among agents does not depend on the fraction of agents in the
alternative states, but rather on the overall number of such agents. Such a
choice makes the transition rates nonextensive, the connectivity between agents
increases with the number of agents $N$. The interactions have a global
character, the range of the correlations involves a macroscopic fraction of
agents. This means that temporal correlations in the level of fluctuations might
be observed for any system size. We will show further that such non-linear terms
in the transition probabilities leads to $1/f$ behavior of the power spectral
density.

The transition probabilities imply the Master equation for the probability
$P_{n}(t)$ to find $n$ agents in the state $1$ at time $t$ \cite{vanKampen1992}:
\begin{multline}
\frac{\partial}{\partial t}P_{n} = p^{+}(n-1)P_{n-1}+p^{-}(n+1)P_{n+1}\\
   -(p^{+}(n)+p^{-}(n))P_{n}\,.\label{eq:master}
\end{multline}
For large enough $N$ we can represent the group dynamics by a continuous
variable $x=n/N$. Using standard methods from ref.~\cite{vanKampen1992}, a
Fokker-Planck equation is derived from the Master equation~(\ref{eq:master})
assuming that $N$ is large and neglecting the terms of the order of $1/N^{2}$:
\begin{multline}
\frac{\partial}{\partial t}P_{x}(x,t)=-\frac{\partial}{\partial x}
h(\varepsilon_{1}(1-x)-\varepsilon_{2}x)P_{x}(x,t) \\
+\frac{1}{2}\frac{\partial^{2}}{\partial x^{2}}
h\left(2x(1-x)+\frac{\varepsilon_{1}}{N}(1-x)
+\frac{\varepsilon_{2}}{N}x\right)P_{x}(x,t)\,,\label{eq:f-p-full}
\end{multline}
where $\varepsilon_1\equiv\sigma_1/h$, $\varepsilon_2\equiv\sigma_2/h$ are
scaled parameters. In the following we will ignore the terms of the order of
$1/N$ in the diffusion term in eq.~(\ref{eq:f-p-full}), assuming that variable
$x$ is not too close to the boundaries $x=0$ and $x=1$, i.e.,
$x\gg\varepsilon_{1}/N$ and $1-x\gg\varepsilon_{2}/N$. In addition we assume
that $\varepsilon_{1},\varepsilon_{2}>0$. Thus the Fokker-Planck equation for
the herding model has the form
\begin{multline}
\frac{\partial}{\partial t}P_{x}(x,t)=-\frac{\partial}{\partial x}
h(\varepsilon_{1}(1-x)-\varepsilon_{2}x)P_{x}(x,t)\\
+\frac{\partial^{2}}{\partial x^{2}}hx(1-x)P_{x}(x,t)\,.\label{eq:f-p}
\end{multline}
This Fokker-Planck equation corresponds to the stochastic differential
equation
\begin{equation}
\upd x=h(\varepsilon_{1}(1-x)-\varepsilon_{2}x)\upd t+\sqrt{2hx(1-x)}\upd W\,,
\label{eq:sde-herding}
\end{equation}
where $W$ is a standard Wiener process (the Brownian motion). The
steady-state solution of eq.~(\ref{eq:f-p}) has the form
\begin{equation}
P_{0}(x)=\frac{\Gamma(\varepsilon_{1}+\varepsilon_{2})}{\Gamma(\varepsilon_{1})
\Gamma(\varepsilon_{2})}x^{\varepsilon_{1}-1}(1-x)^{\varepsilon_{2}-1}\,.
\label{eq:steady}
\end{equation}
Eqs.~(\ref{eq:f-p})--(\ref{eq:steady}) were obtained in ref.~\cite{Alfarano2005}.

Using $\varepsilon_{1}=\varepsilon_{2}$ in eq.~(\ref{eq:steady})
we recover the equilibrium distribution as in ref.~\cite{Kirman1993}.
The distribution (\ref{eq:steady}) exhibits a unique mode if both
parameters take a value larger than $1$, while it shows bi-modality
for the case $\varepsilon_{1},\varepsilon_{2}<1$. Furthermore, the
distribution shows a monotonic behavior if one parameter is larger
than $1$ and the other smaller than $1$.

\section{Nonlinear stochastic differential equation generating signals with
$1/f^{\beta}$ noise}

\label{sec:sde}Starting from the point process model, proposed and analyzed in
Refs.~\cite{Kaulakys1998-1,Kaulakys1999-1,Kaulakys2000-1,
Kaulakys2000-2,Gontis2004,Kaulakys2005}, the nonlinear SDEs generating
processes with $1/f^{\beta}$ noise are derived
\cite{Kaulakys2004,Kaulakys2006,Kaulakys2009}. The general expression for the
proposed class of It\^o SDEs is
\begin{equation}
\upd x=\sigma^{2}\left(\eta-\frac{1}{2}\lambda\right)x^{2\eta-1}\upd t
+\sigma x^{\eta}\upd W\,.\label{eq:sde}
\end{equation}
Here $x$ is the signal, $\eta\neq1$ is the power-law exponent of the
multiplicative noise, $\lambda$ defines the behavior of stationary probability
distribution, and $W$ is a standard Wiener process (the Brownian motion). The
Fokker-Planck equation corresponding to SDE~(\ref{eq:sde}) gives the power-law
probability density function (PDF) of the signal intensity $P_{0}(x)\sim
x^{-\lambda}$ with the exponent $\lambda$. In
Refs.~\cite{Kaulakys2005,Kaulakys2006} it was shown that SDE~(\ref{eq:sde})
generates signals with power spectral density
\begin{equation}
S(f)\sim \frac{1}{f^{\beta}} \,,\qquad\beta=1+\frac{\lambda-3}{2(\eta-1)}\,.\label{eq:beta}
\end{equation}
The nonlinear SDE~(\ref{eq:sde}) has the simplest form of the multiplicative
noise term, $\sigma x^{\eta}dW$. Multiplicative equations with the drift
coefficient proportional to the Stratonovich drift correction for transformation
from the Stratonovich to the It\^o stochastic equation \cite{Arnold2000}
generate signals with the power-law distributions \cite{Kaulakys2009}.
Equation~(\ref{eq:sde}) is of such a type and has the stationary PDF of the
power-law form. The connection of the power spectral density of the signal
generated by SDE~(\ref{eq:sde}) with the behavior of the eigenvalues of the
corresponding Fokker-Planck equation was analyzed in ref.~\cite{Ruseckas2010}.
In ref.~\cite{Erland2011} it is shown that $1/f^{\beta}$ noise is
equivalent to a Markovian eigenstructure power relation.

SDE (\ref{eq:sde}) exhibits the following scaling property: changing the
stochastic variable from $x$ to a scaled variable $x^{\prime}=ax$ changes the
time-scale of the equation to $t^{\prime}=a^{2(1-\eta)}t$, leaving the from of
the equation unchanged. This scaling property is one of the reasons for the
appearance of the $1/f^{\beta}$ power spectral density.

Another remarkable property of SDE~(\ref{eq:sde}) is the behavior under
transformation of the variable $x$: if instead of $x$ we introduce
\begin{equation}
y=x^{\alpha}\,,\label{eq:transform1}
\end{equation}
then from eq.~(\ref{eq:sde}) we get the equation of the same type
\begin{equation}
\upd y=\sigma^{\prime 2}\left(\eta^{\prime}
-\frac{\lambda^{\prime}}{2}\right)y^{2\eta^{\prime}-1}\upd t
+\sigma^{\prime}y^{\eta^{\prime}}\upd W\,,
\end{equation}
only with different parameters $\sigma^{\prime} = \alpha\sigma$,
$\eta^{\prime} = (\eta-1)/\alpha + 1$,
$\lambda^{\prime} = (\lambda-1)/\alpha + 1$.

For $\lambda>1$ the distribution $P_{0}(x)$ diverges as $x\rightarrow0$,
therefore the diffusion of stochastic variable $x$ should be restricted at least
from the side of small values, or equation (\ref{eq:sde}) should be modified.
The simplest choice of the restriction is the reflective boundary conditions at
$x=x_{\mathrm{min}}$ and $x=x_{\mathrm{max}}$. However, other forms of
restrictions are possible and have been considered, as well. Exponentially
restricted diffusion is generated by the SDE
\begin{multline}
\upd x=\sigma^{2}\left[\eta-\frac{1}{2}\lambda+\frac{m}{2}
\left(\frac{x_{\mathrm{min}}^{m}}{x^{m}}
-\frac{x^{m}}{x_{\mathrm{max}}^{m}}\right)\right]x^{2\eta-1}\upd t\\
+\sigma x^{\eta}\upd W
\label{eq:sde-restricted}
\end{multline}
obtained from eq.~(\ref{eq:sde}) by introducing the additional terms.

For $\lambda=3$ we get that $\beta=1$ and SDE (\ref{eq:sde}) gives signal
exhibiting $1/f$ noise. Numerical solution of the equations confirms the
presence of the frequency region for which the power spectral density has
$1/f^{\beta}$ dependence. The width of this region can be increased by
increasing the ratio between the minimum and the maximum values of the
stochastic variable $x$. In addition, the region in the power spectral density
with the power-law behavior depends on the exponent $\eta$: if $\eta=1$ then
this width is zero; the width increases with increasing the difference
$|\eta-1|$ \cite{Ruseckas2010}.

For some choices of parameters, SDE~(\ref{eq:sde}) or its variant
(\ref{eq:sde-restricted}) takes the form of the well-known SDEs considered in
econopysics and finance. In case when the exponent of multiplicative noise
$\eta=0$ and $\sigma=1$, (\ref{eq:sde}) takes the form of the SDE for the
\emph{Bessel process} \cite{Jeanblanc2009},
\begin{equation}
\upd x=\frac{\delta-1}{2}\frac{1}{x}\upd t+\upd W\,,
\end{equation}
of dimension $\delta=1-\lambda$, while $\eta=1/2$ and $\sigma=2$
corresponds to the \emph{squared Bessel process} \cite{Jeanblanc2009},
\begin{equation}
\upd x=\delta \upd t+2\sqrt{x}\upd W\,,
\end{equation}
of dimension $\delta=2(1-\lambda)$. SDE with exponential restriction
(\ref{eq:sde-restricted}) for $\eta=1/2$, $x_{\mathrm{min}}=0$
and $m=1$ gives the \emph{Cox-Ingersoll-Ross} (CIR) process \cite{Jeanblanc2009},
\begin{equation}
\upd x=k(\theta-x)\upd t+\sigma\sqrt{x}\upd W\,,
\end{equation}
where $k=\sigma^{2}/2x_{\mathrm{max}}$ and $\theta=x_{\mathrm{max}}(1-\lambda)$.
When $\nu=2\eta$, $x_{\mathrm{max}}=\infty$ and $m=2\eta-2$ then
eq.~(\ref{eq:sde-restricted}) takes the form of the \emph{Constant Elasticity
of Variance} (CEV) process \cite{Jeanblanc2009},
\begin{equation}
\upd x=\mu x\upd t+\sigma x^{\eta}\upd W\,,
\end{equation}
where $\mu=\sigma^{2}(\eta-1)x_{\mathrm{min}}^{2(\eta-1)}$.

The numerical analysis of the proposed SDE~(\ref{eq:sde}) reveals the secondary
structure of the signal composed of peaks or bursts, corresponding to the large
deviations of the variable $x$ from the proper average fluctuations. Bursts are
characterized by power-law distributions of burst size, burst duration, and
interburst time \cite{Kaulakys2009}. Therefore, proposed nonlinear SDEs may
simulate avalanches in self-organized critical (SOC) models and extreme event
return times in long memory processes.

\section{Herding model and $1/f$ noise}

\label{sec:herding-1f}Let us consider the case when $x\ll1$. Then
SDE~(\ref{eq:sde-herding}) approximately has the form
\begin{equation}
\upd x\approx h(\varepsilon_{1}-\varepsilon_{2}x)\upd t
+\sqrt{2hx}\upd W\,.\label{eq:sde-herding-approx}
\end{equation}
Eq.~(\ref{eq:sde-herding-approx}) has the form of our nonlinear
SDE~(\ref{eq:sde}) with the multiplicative noise and having parameters $\eta =
1/2$, $\lambda = 1-\varepsilon_1$. The term with $\varepsilon_{2}$ in
eq.~(\ref{eq:sde-herding-approx}) gives restrictions at larger $x$. The possible
values of the parameter $\varepsilon_{1}$ are restricted to $\varepsilon_{1}>0$
and this limits the possible values of the exponents $\lambda$ and $\beta$. In
particular, it is not possible to obtain $1/f$ noise with $\beta=1$
for the herding dynamics of population $x$. Nevertheless,
the nonlinear form of Eq.~(\ref{eq:sde-herding-approx}) allows to apply
transformation (\ref{eq:transform1}) and get a flexible choice of variables
and corresponding exponents $\lambda$ and $\beta$.

One possible choice is to consider $y=1/x$, having $\alpha=-1$ in
eq.~(\ref{eq:transform1}). The range of possible values of $1/x$ is
$[1,+\infty)$. Since only large values of $y$ are relevant for obtaining $1/f$
noise, this range can be extended to include zero by introducing
$y=1/x-1=(1-x)/x$. This variable $y$ has a clear interpretation: it is equal to
the ratio of the number of agents in the state $2$ to the number of agents in
the state $1$:
\begin{equation}
y=\frac{1-x}{x}=\frac{N-n}{n}\,.\label{eq:y-def-n}
\end{equation}
A stochastic variable similar to $y$ (\ref{eq:y-def-n}) was used in
ref.~\cite{Alfarano2005} to model absolute return, while our variable $x$
corresponds to a fraction of fundamentalists in ref.~\cite{Alfarano2005}.
Transformation (\ref{eq:y-def-n}) of variables leads from
SDE~(\ref{eq:sde-herding}) to
\begin{equation}
\upd y=h[(2-\varepsilon_{1})y+\varepsilon_{2}](1+y)\upd t
+\sqrt{2hy}(1+y)\upd W\,.\label{eq:sde-y}
\end{equation}
Similar equation has been obtained in ref.~\cite{Kononovicius2011}.
Steady state PDF of the new variable $y$ is
\begin{equation}
P_{0}(y)=\frac{\Gamma(\varepsilon_{1}+\varepsilon_{2})}{\Gamma(\varepsilon_{2})\Gamma(\varepsilon_{1})}\frac{y^{\varepsilon_{2}-1}}{(1+y)^{\varepsilon_{2}+\varepsilon_{1}}}\,.\label{eq:steady-y}
\end{equation}
When $y\gg1$ then eq.~(\ref{eq:sde-y}) approximately has the form
\begin{equation}
\upd y\approx h(2-\varepsilon_1)y^2\upd t
+\sqrt{2h}y^{\frac{3}{2}}\upd W\,. \label{eq:sde-y-approx}
\end{equation}
Eq.~(\ref{eq:sde-y-approx}) has the form of our nonlinear SDE~(\ref{eq:sde})
with multiplicative noise, having parameters $\eta = 3/2$, $\lambda =
1+\varepsilon_1$. Eq.~(\ref{eq:sde-y-approx}) is similar to a well known $3/2$
model of stochastic volatility \cite{Ahn1999}. According to eq.~(\ref{eq:beta}),
the power exponent of the power spectral density is $\beta=\varepsilon_1 - 1$.
If $\varepsilon_1 = 2$, we obtain $1/f$ spectrum. Thus it is possible to obtain
a signal with $1/f$ noise starting form the herding model. SDE~(\ref{eq:sde-y})
demonstrates yet another possible form of the restriction of diffusion of the
stochastic variable in SDE~(\ref{eq:sde}) from the side of small values.
It is of interest to note, that the strong herding tendency for
$h>\sigma_1 /2$ ($\epsilon_1<2$) yields the long-range process with $\beta<1$,
the power-law correlation $C(t) \sim 1/t^{1-\beta}$, and
distribution~(\ref{eq:steady-y}) with the diverging variance. 

\begin{figure}
\includegraphics[width=0.45\textwidth]{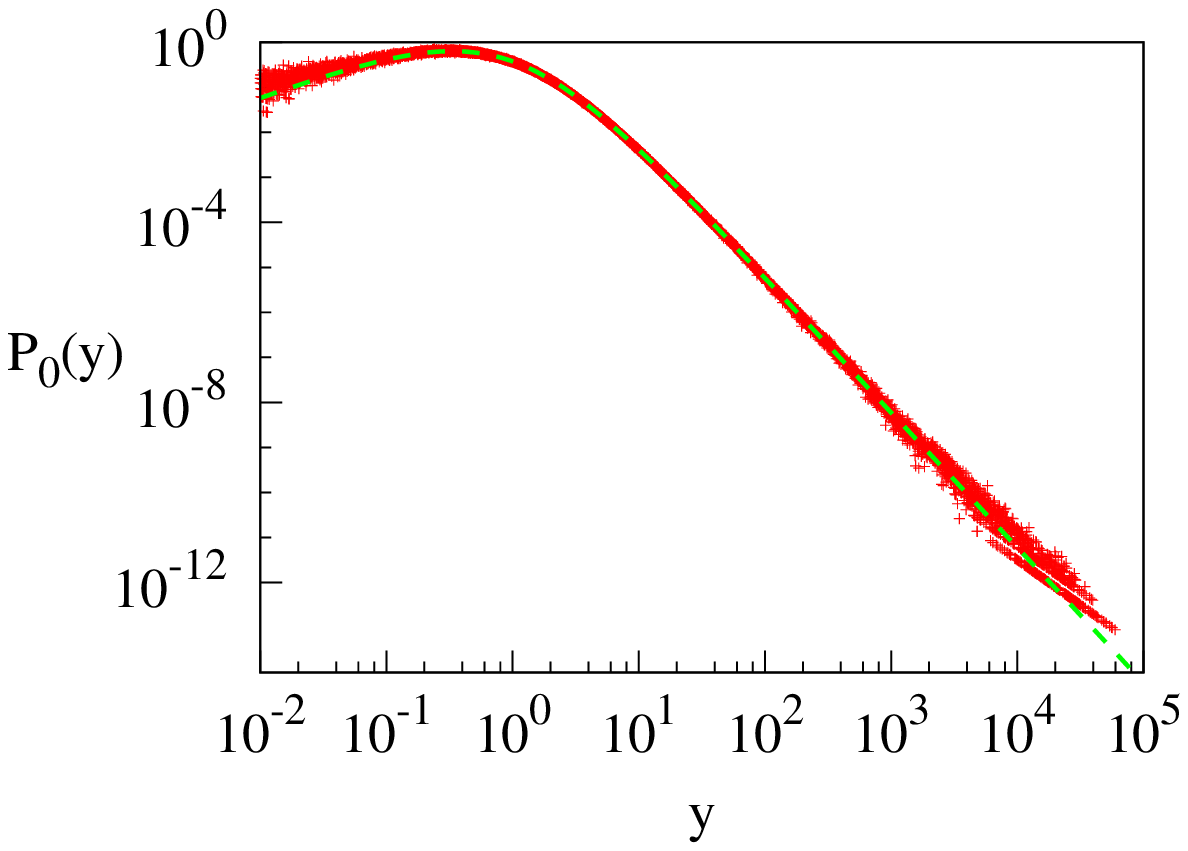}
\includegraphics[width=0.45\textwidth]{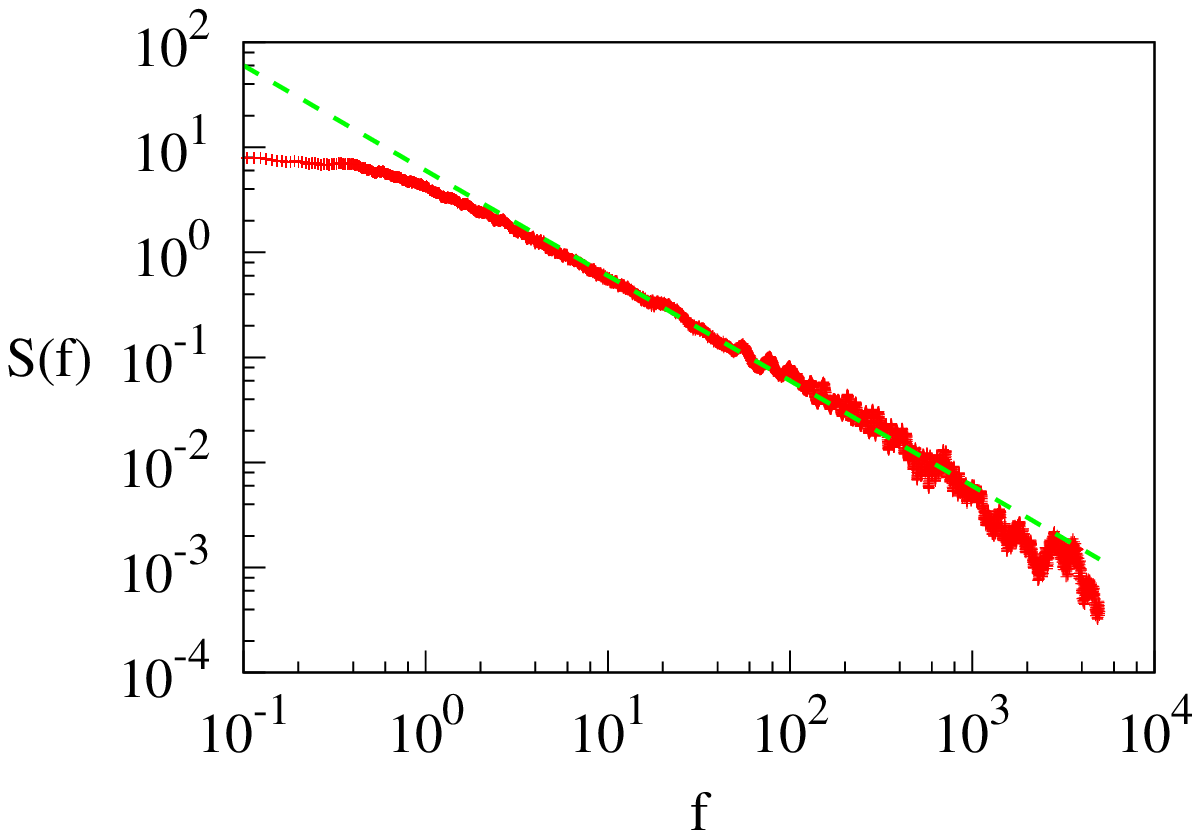}
\caption{(Color online) Steady state PDF $P_0(y)$ (upper panel) and power
spectral density $S(f)$ (lower panel) of the signal generated by
eq.~(\ref{eq:sde-y}). The dashed (green) lines are the analytical expression
(\ref{eq:steady-y}) for the steady state PDF (upper panel) and the
slope $1/f$ (lower panel). The parameters used are
$\varepsilon_{1}=\varepsilon_{2}=2$ and $h=1$.}
\label{fig:sde-y}
\end{figure}

Comparison of numerically obtained steady state PDF and power spectral
density of the signal generated by eq.~(\ref{eq:sde-y}) with analytical
expressions is presented in fig.~\ref{fig:sde-y}. For the numerical
solution we use Euler-Marujama approximation with variable step of
integration, transforming the differential equations to the difference
equations \cite{Kaulakys2004,Kaulakys2006}. We see a good agreement
of the numerical results with the analytical expressions. Numerical
solution of the equations confirms the presence of the frequency region
for which the power spectral density has $1/f$ dependence.

\begin{figure}
\includegraphics[width=0.45\textwidth]{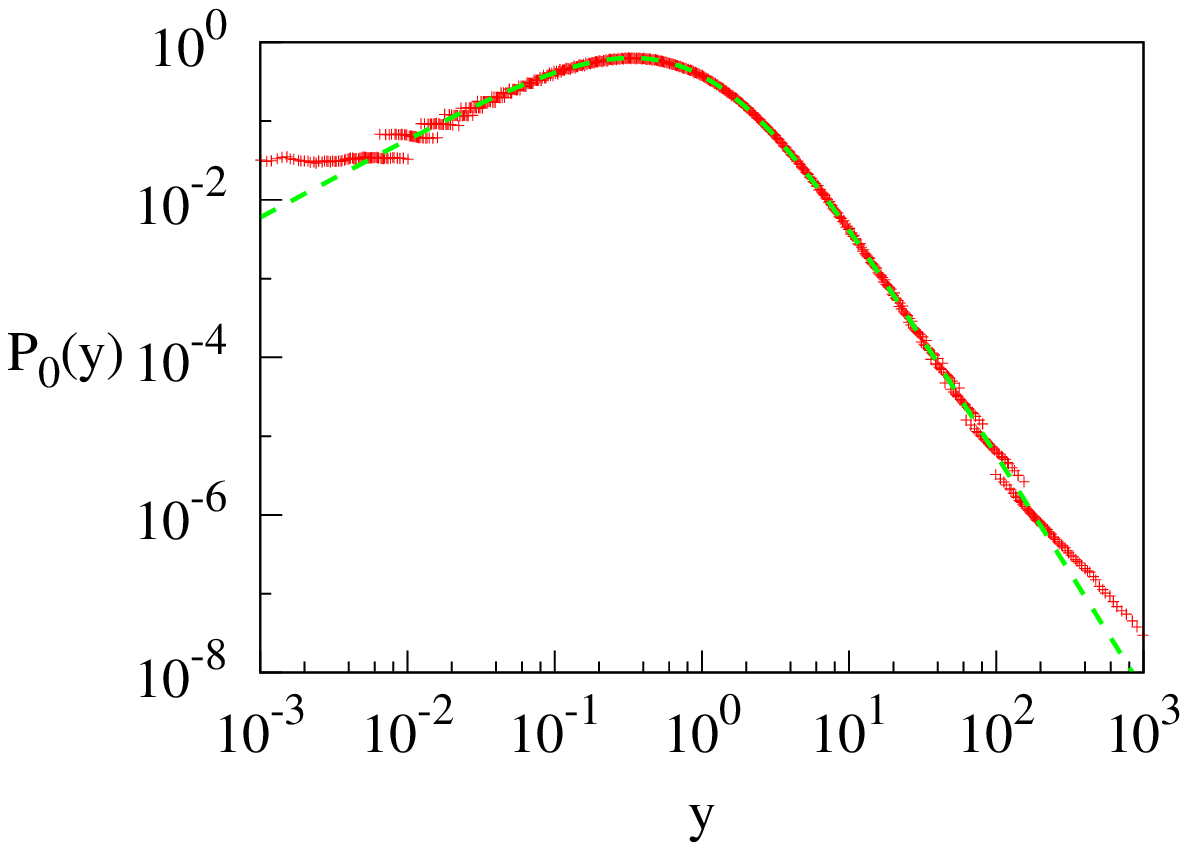}
\includegraphics[width=0.45\textwidth]{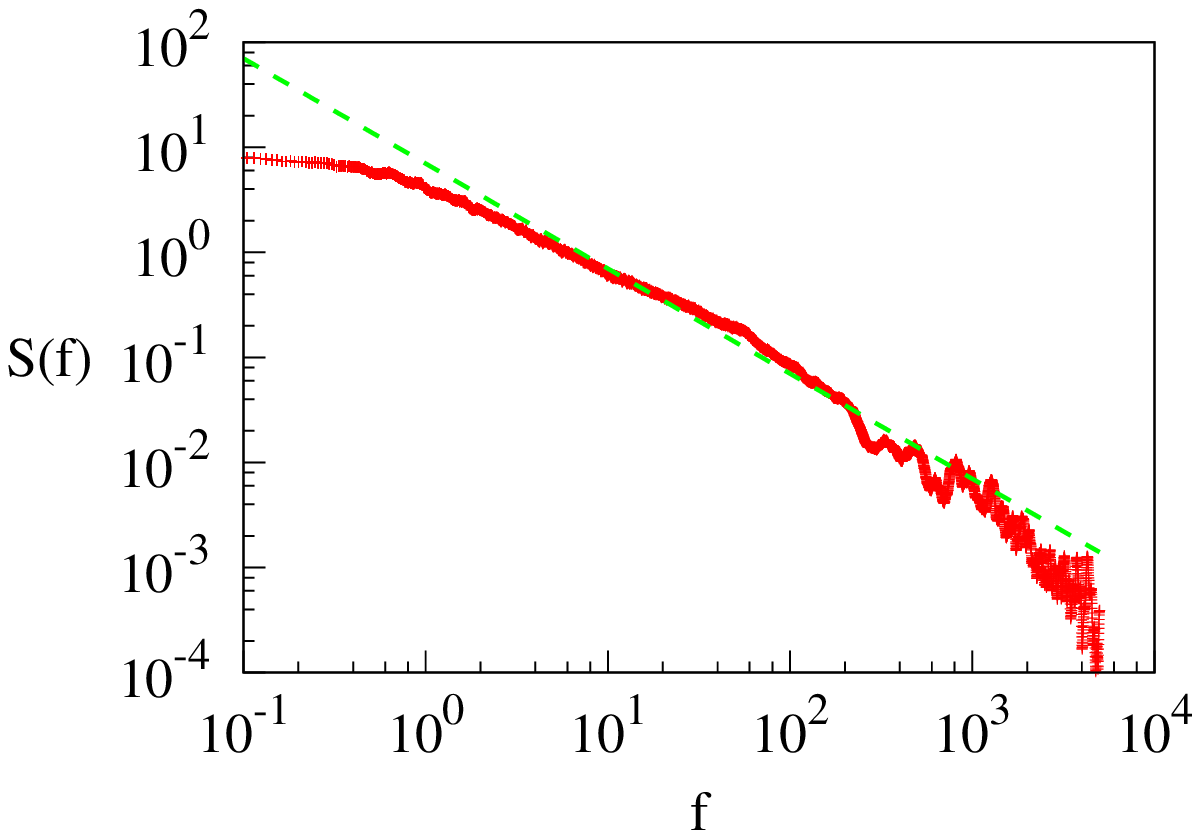}
\caption{(Color online) Steady state PDF $P_0(y)$ (upper panel) and power
spectral density $S(f)$ (lower panel) of the stochastic variable $y$ calculated
according to eq.~(\ref{eq:y-def-n}), using the number of agents $n$ obtained
from the jump process defined by the transition probabilities (\ref{eq:pp}) and
(\ref{eq:pm}). The dashed (green) lines are the analytical expression
(\ref{eq:steady-y}) for the steady state PDF (upper panel) and the
slope $1/f$ (lower panel). The total number of agents is $N=10000$,
other parameters are the same as in fig.~\ref{fig:sde-y}.}
\label{fig:agent}
\end{figure}

For the comparison, steady state PDF and power spectral density of the
stochastic variable $y$ calculated using the number of agents $n$ according to
eq.~(\ref{eq:y-def-n}), is presented if fig.~\ref{fig:agent}. The number of
agents $n$ is obtained from the jump process defined by the transition
probabilities (\ref{eq:pp}) and (\ref{eq:pm}). We see a good agreement of the
numerical results with the analytical expressions. Thus, numerical calculations
confirm the SDE~(\ref{fig:sde-y}) and the possibility to obtain $1/f$ spectrum
from a herding model.

\section{Possible generalizations}

\label{sec:general}One of the possible generalizations of the model
presented above is to consider a stochastic variable $y$ defined
as
\begin{equation}
y=\left(\frac{1-x}{x}\right)^{1/\alpha}\,.
\end{equation}
Then the transformation of variables leads from SDE~(\ref{eq:sde-herding})
to the SDE
\begin{multline}
\upd y=\frac{h}{\alpha}\left[\left(1+{\textstyle\frac{1}{\alpha}}
-\varepsilon_1\right)
+\left(\varepsilon_2+{\textstyle\frac{1}{\alpha}}-1\right)y^{-\alpha}\right]
y(1+y^{\alpha})\upd t\\
+\frac{\sqrt{2h}}{\alpha}y^{1-\frac{\alpha}{2}}(1+y^{\alpha})\upd W\,.
\end{multline}
The corresponding steady state PDF is
\begin{equation}
P_{0}(y)=\frac{\alpha\Gamma(\varepsilon_{1}+\varepsilon_{2})}{\Gamma(\varepsilon_{2})\Gamma(\varepsilon_{1})}\frac{y^{\alpha\varepsilon_{2}-1}}{(1+y^{\alpha})^{\varepsilon_{2}+\varepsilon_{1}}}\,.\label{eq:pdf-y}
\end{equation}
It should be noted that PDF~(\ref{eq:pdf-y}) for some choice of parameters has
the form of distributions featured in nonextensive statistical mechanics
\cite{Tsallis1988,Tsallis1998,Prato1999,Tsallis1999,Tsallis2009-1,Tsallis2009-2,Tsallis2011}:
the values of the parameters $\alpha=1$, $\varepsilon_{2}=1$ correspond to
$q$-exponential distribution with $q=1+1/(1+\varepsilon_{1})$ and the values of
the parameters $\alpha=2$, $\varepsilon_{2}=1/2$ correspond to $q$-Gaussian
distribution with $q=1+2/(1+2\varepsilon_{1})$. When $y\gg1$ then we get
SDE~(\ref{eq:sde}) with parameters $\eta = 1 + \alpha/2$, $\lambda =
1+\alpha\varepsilon_1$. According to eq.~(\ref{eq:beta}), the power exponent of
the power spectral density is $\beta=\varepsilon_1 + 1 - 2/\alpha$.

Another way to generalize the herding model is to introduce the additional
non-linearities into transition probabilities (\ref{eq:pp}) and (\ref{eq:pm}).
The original transition probabilities (\ref{eq:pp}) and (\ref{eq:pm})
assume that agents meet at a constant rate and therefore the coefficients
$\sigma_{1}$, $\sigma_{2}$ and $h$ are constant. One possibility
to extend the model is to assume that the rate at which the agents
meet depends on the global state of the system. In such a situation
the new transition probabilities can be written as
\begin{align}
p(n\rightarrow n+1) & = \frac{1}{\tau(n)}(N-n)(\sigma_{1}+hn)\,,\label{eq:pp-1}\\
p(n\rightarrow n-1) & = \frac{1}{\tau(n)}n(\sigma_{2}+h(N-n))\,,\label{eq:pm-1}
\end{align}
where $\tau(n)$ describes the time scale of the microscopic events.
Assuming that the time scale $\tau$ depends only on the ratio $x=n/N$,
the SDE obtained form the modified model instead of eq.~(\ref{eq:sde-herding})
has the form
\begin{equation}
\upd x=\frac{h}{\tau(x)}(\varepsilon_{1}(1-x)-\varepsilon_{2}x)\upd t
+\sqrt{2\frac{h}{\tau(x)}x(1-x)}\upd W\,,
\end{equation}
whereas the SDE for the variable $y=(1-x)/x$ is
\begin{equation}
\upd y=\frac{h}{\tau_{y}(y)}[(2-\varepsilon_{1})y+\varepsilon_{2}](1+y)\upd t
+\sqrt{2\frac{h}{\tau_{y}(y)}y}(1+y)\upd W\,.\label{eq:sde-y-mod}
\end{equation}
Here $\tau_{y}(y)\equiv\tau(1/(1+y))$. Similar modification of the
herding model has been proposed in ref.~\cite{Kononovicius2011}.

Let us consider the case of $\tau(y)=y^{-\gamma}$. Then eq.~(\ref{eq:sde-y-mod})
becomes
\begin{equation}
\upd y=h[(2-\varepsilon_{1})y+\varepsilon_{2}]y^{\gamma}(1+y)\upd t
+\sqrt{2hy^{1+\gamma}}(1+y)\upd W\,.
\end{equation}
In the limit $y\gg1$ we get SDE~(\ref{eq:sde}) with parameters
$\eta = 3/2+\gamma/2$, $\lambda = \varepsilon_1 + 1 + \gamma$.
According to eq.~(\ref{eq:beta}), the power exponent of the power
spectral density is
\begin{equation}
\beta=1+\frac{\lambda-3}{2(\eta-1)}=1+\frac{\varepsilon_{1}+\gamma-2}{1+\gamma} \,.
\end{equation}
Thus we have shown that it is possible to obtain different values
of the parameter $\eta$ in eq.~(\ref{eq:sde}).

\section{Conclusions}

\label{sec:concl}Starting from a simple agent-based model describing herding
behavior we obtained non-linear SDE for the agent population $x$
equal to the fraction of agents in the state $1$. For $x\ll1$ this non-linear
SDE has the form similar to SDE proposed in
Refs.~\cite{Kaulakys2004,Kaulakys2006} for the modeling of $1/f$ noise. This
form suggests that it might be possible with the appropriate
transformation of variables to obtain signals having $1/f$ behavior of the
power spectral density from this agent model. However, the possible values of
the parameter $\varepsilon_{1}$ in the model,
eq.~(\ref{eq:sde-herding-approx}), are restricted to the positive
values and this limits the values of the exponents $\lambda$ and
$\beta$ of the power-law statistics. The solution is to consider not the
fraction of agents $x$, but another variable $y$, equal to the ratio of number
of agents in the state $2$ to the number of agents in the state $1$. This new
variable is related to $x$ by a simple transformation. The non-linear SDE for
the stochastic variable $y$ in the limit of large values $y\gg1$ has the
required form for obtaining $1/f$ noise. This result shows that it is possible
to obtain $1/f$ noise starting from the agent-based herding model and
introducing the appropriate variables. These analytical results are checked by
numerical calculations, showing good agreement with analytical predictions.
Thus, herding dynamics can be seen as a microscopic explanation of
proposed non-linear SDEs generating signals with $1/f^\beta$ spectrum. The
derivation of SDEs shows that non-linear terms in the transition probabilities,
describing global interactions between agents, are crucial to the appearance of
$1/f$ noise. The exponents $\lambda$ and $\beta$ of the power-law
statistics can be adjusted by introducing feedback between the macroscopic
system state $x$ and the rate of the microscopic events $1/\tau(x)$.
Application of the similar model for description of the return in the
finacial markets has been proposed in ref.~\cite{Kononovicius2011}.

\end{document}